\documentclass[amsmath,amssymb,preprint,showpacs,preprintnumbers,floatfix,aps,pra]{revtex4-1}
\usepackage{graphicx}
\usepackage{natbib}
\usepackage{dcolumn}
\usepackage{bm}
\usepackage{amsmath}
\usepackage{array}
\usepackage[utf8]{inputenc}
\usepackage{hyperref}
\usepackage{balance}
\usepackage{multirow}
\usepackage{longtable}
\usepackage{epstopdf}
\usepackage{comment}
\usepackage{xcolor}

\usepackage{dcolumn}
\renewcommand{\k}{{\bbox k}}
\renewcommand{\k}{{\bm k}}
\newcommand{\hs}{\hspace*}

\newcommand{\Eref}[1] {Eq.~(\ref{#1})}

\newcommand{\nn}{\nonumber}
\newcommand{\be}{\begin{equation}}
\newcommand{\ee}{\end{equation}}
\newcommand{\br}{\begin{eqnarray*}}
\newcommand{\er}{\end{eqnarray*}}
\newcommand{\ba}{\begin{eqnarray}}
\newcommand{\ea}{\end{eqnarray}}
\newcommand{\bp}{\begin{minipage}}
\newcommand{\ep}{\end{minipage}}

\newcommand{\bt}{\begin{tabular}}
\newcommand{\et}{\end{tabular}}

\renewcommand{\r}{{\bbox r}}
\renewcommand{\r}{{\bm r}}

\newcommand{\z}{{\bbox z}}
\renewcommand{\z}{{\bm z}}

\begin{document}

\title{Crossover between the zeptosecond and attosecond physics }
\author{T. Nandi$^1$, Yash Kumar$^2$, Adya P. Mishra$^{3,4}$, Nishchal R. Dwivedi$^{5}$, Chandra Kumar$^6$, Gajendra Singh$^7$, N.Sowmya$^8$, H.C.Manjunatha$^9$, Sudhir R. Jain$^{11}$, and A.S. Kheifets$^{12}$}
\affiliation{$^{1}$ Department of Physics, Ramakrishna Mission Vivekananda Educational and Research Institute, PO Belur Math, Dist Howrah 711202, West Bengal, India}
\affiliation{$^2$ Université Bourgogne Europe, Dijon, France}
\affiliation{$^3$Atomic \& Molecular Physics Division, Bhabha Atomic Research Centre, Trombay, Mumbai - 400 085, India}
\affiliation{$^4$Homi Bhabha National Institute, Training School Complex, Anushakti Nagar, Mumbai - 400 094, India}
\affiliation{$^5$Department of Basic Science and Humanities, SVKM’s NMIMS Mukesh Patel School of Technology Management \& Engineering, Mumbai -400 056, India}
\affiliation{$^6$ Inter-University Accelerator Centre, JNU New Campus, Aruna Asaf Ali Marg,  New Delhi-110067, India.}
\affiliation{$^7$Institute for Plasma Research, Bhat 382 428, Gandhinagar, India}
\affiliation{$^8$Department of Physics, Government First Grade College, Chikkaballapur-562101, Karnataka, India}
\affiliation{$^9$Department of Physics, Government First Grade College, Devanahalli-562110, Karnataka, India}
\affiliation{$^{10}$Nuclear Physics Division, Bhabha Atomic Research Centre, Trombay, Mumbai - 400 085, India}
\affiliation{ $^9$Department of Physics, Kalyani University, Kalyani, Nadia, West Bengal-741235, India}
\affiliation{$^{11}$UM-DAE Centre for Excellence in Basic Sciences, University of Mumbai, Vidyanagari Campus, Mumbai - 400 098, India}
\affiliation{$^{12}$Research School of Physics and Engineering, The Australian National University, Canberra, Australian Capital Territory 0200, Australia}
\newpage
\begin{abstract}
\textbf{Nuclear orbiting resonances have been revealed at the sub-barrier energies as an atomic phenomenon by means of x-ray spectroscopy experiments. This interpretation is supported by several phenomenological models and theoretical estimates of the nuclear orbiting timescale and cross-section, inelastic scattering cross section including both nuclear and Coulomb excitation, and the Wigner-Smith time delay. We demonstrate that a multi-photon exchange during nuclear orbiting is responsible for an atomic excitation. Furthermore, proximity of the projectile and target nucleus during the nuclear orbiting modifies the effective charge of the projectile. Even though this orbiting induced excitation is triggered in zeptoseconds,  it can still be observed in the attosecond time scale because of the Wigner-Smith time delay inherent to autoionization. Thus, we demonstrate the crossover between the zeptosecond and attosecond time scales which are native to nuclear and atomic physics, respectively. Markedly, this crossover may be the reason for x-ray production from ultra short nuclear processes ($\leq  10^{-21}$ sec). This explanation is likely to resolve the fission time scale anomaly and can stimulate cross-disciplinary research ranging from solid state to high-energy physics.}
\end{abstract}


\maketitle

\section{Introduction}
Despite significantly  large differences between the range and coupling constant of the electromagnetic and strong forces, the atomic and nuclear 
phenomena interfere with each other giving rise to various exotic events \cite{Freedman-ARNS-1974}. 
One example of such events has been discovered recently by  Sharma and Nandi \cite{Prashant-PRL-2017} who have observed unusual resonance-like structures in the x-ray spectra with the beam energy approaching the fusion barrier energy. We call this particular position the resonance energy ($B_{res}$) appearing between the interaction barrier ($B_{int}$, the threshold barrier from where nuclear interaction begins) and  fusion barrier ($B_{fus}$) energies \cite{BASS197445}. The observed resonance structures in the projectile x-ray spectra, as shown in Fig. 2 of  ref. \cite{Prashant-PRL-2017}, have been attributed to the shake off ionization due to sudden nuclear recoil while the projectile is moving on the Rutherford scattering trajectory. Even though quite good agreement has been found between theory and experiment  \cite{Prashant-PRL-2017} for variation of mean charge state of the projectile ions versus beam energy \cite{Prashant-PLA}, certain facts still remain unexplained, viz., (i) unusually high scattering cross-section at large angles (anomalous elastic scattering) in light-heavy ion collisions $(20 \leq (A_{T} + A_{P}) \leq 80)$  \cite{Braun}, and (ii) the resonance-like structure observed near the fusion barrier instead of the interaction barrier \cite{Prashant-PRL-2017}.
\par
To explain the shortcomings mentioned above, we consider that besides the usual elastic scattering, some other process must be playing important roles in the observed resonance phenomenon. We have identified a process called nuclear orbiting resonance (dinuclear complexes) to explain every issue associated with the observed resonance. However, it gives rise to a fundamental puzzling question as to how a nuclear phenomenon is being observed through an atomic process. It is well known that every quantum system  evolves on its characteristic time scale which varies widely between solids (picoseconds \cite{Haller2013}), molecules (femtoseconds  \cite{PNAS}), atoms (attoseconds  \cite{PRL2001}), nuclei (zeptoseconds  \cite{PRL2002}), and quarks (yoctoseconds  \cite{PRL2009}). Though a crossover between these time scales is very rare in nature, a crossover between attophysics and femtochemistry due to  ultrafast hole migration has been discovered \cite{NatPhotonics2014}. In this article, we reveal that the nuclear orbiting phenomenon crosses the boundary between the nuclear  and atomic  time scales because of Wigner-Smith time delay \cite{wigner55,smith60} and thus it allows us to discover a crossover between zeptophysics and attophysics. 
\section{Nuclear phenomena which occurred in zeptosecond time scales, but manifested in attosecond time scales}
Very often various ultra fast ($\leq  10^{-21}$ sec) nuclear phenomena such as deep inelastic collisions \cite{DIC}, multi nucleon transfer reaction \cite{Nandi2009}, internal conversion \cite{IC}, nuclear recoil induced shake off ionization \cite{Prashant-PRL-2017} give rise to slow process occurring in $\ge 10^{-18}$ sec in terms of  x-ray emissions. Even though x-ray spectroscopy has been used in numerous measurements to study these nuclear processes, no one has asked yet the fundamental intriguing question how a phenomenon in zeptosecond is being manifested in attosecond time scale. 
\par
Let us first consider nuclear processes that do not involve any change in atomic number of the reaction partners. Inelastic scattering process including nuclear and Coulomb excitation in projectile as well as target nuclei is  one of such processes, which takes place through photon exchange between the target and projectile nuclei. In inelastic scattering \cite{thompson1988coupled}, a part of the kinetic energy of the system is used to activate the nucleus into an excited state. Calculation of the corresponding cross sections are discussed in following section. Neutron transfer reaction is another possibility. However, the ground state {\textit{Q}}-values for one neutron transfer process for the reactions considered in ref. \cite{Prashant-PRL-2017} ($^{12}$C($^{56}$Fe,$^{56}$Fe), $^{12}$C($^{58}$Ni,$^{58}$Ni) and $^{12}$C($^{63}$Cu,$^{63}$Cu)) lie in the range of -6 to -11 MeV; therefore, the neutron transfer is hardly possible in the sub-barrier energy regime. Next, to examine the relative position of the $B_{res}$ accurately, a thorough study involving many semiempirical calculations reported recently \cite{deepak} shows that the Broglia and Winther (BW) model \cite{BW} is the most accurate for the fusion barrier, whereas the model  \cite{deepak} is the best for the interaction barrier. Results given in Table \ref{first} show that the $B_{res}$ lies closer to $B_{fus}$.
\begin{table}
\centering
\caption{The calculated values of angular momentum (rounded off) and orbiting radius, $R_o$, of the dinuclear systems formed at the resonance energies. The centre of mass energies for $B_{res}$ and $B_{int}$ are in MeV. $B_{res}$ is the measured value of ref. \cite{Prashant-PRL-2017} and $B_{int}$ is our calculated values \cite{deepak}. The radius $R_{t}$ (touching radius is sum of radii of projectile and target nuclei), $R_b $ (Fusion barrier radius), and $R_o$ are in fm. $R_{t}$ and $R_b $ are taken from the BW model \cite{BW}.}
\begin{tabular}{c c c c c c c c}
\hline
$System$ & $B_{res}$ & $B_{int}$ & $B_{fus}$ & $R_{t}$ & $R_b $ & $R_o$ & $L (\hbar)$\\
\hline
$^{12}C$+$^{56}Fe$ & $21.17$ &$17.43$ &$22.68$ & $7.14$ &$9.23$ &$9.02$ &$12$ \\
$^{12}C$+$^{58}Ni$ &$22.97$ &$19.48$&$24.37$ &$7.20$ &$9.25$&$9.06$ &$12$ \\
$^{12}C$+$^{63}Cu$ &$22.88$ &$20.05$ &$24.89$ & $7.34$ & $9.39$ & $9.23$ & $11$ \\
\hline
\label{first}
\end{tabular}
\end{table}
\par
We now consider whether an elastic process can be suitable to explain the observed resonance structure  \cite{Prashant-PRL-2017}. In 1970s, it was noticed for the light-heavy ion systems ($ 20 \leq (A_{T}+A_{P}) \leq 80 $) that the measured elastic cross-section was higher in the large-angle region ($100{^\circ} - 180^{\circ}$) than that of the optical model predictions \cite{Braun}. This anomalous behavior was attributed to a process, called nuclear orbiting resonance, in which the projectile nucleus goes around the target nucleus for a long-while as a dinuclear complex before it  exits through the entrance channel  \cite{Shapira,Bertulani,Wheeler,SANDERS1999487}. It thus provides us with an indication that the nuclear orbiting resonance can be responsible for the observed resonance \cite{Prashant-PRL-2017}, and we proceed to examine this possibility.
One necessary condition for nuclear orbiting  \cite{Shapira} is that there must exist a pocket in the total nucleus-nucleus potential due to nuclear, Coulomb, and centrifugal forces, such that
\begin{equation}
\frac{d(V(r)}{dr}\bigg|_{r=R_{o}} = 
\frac{d(V_{nucl}(r)+V_{coul}(r)+V_{cent}(r))}{dr}\bigg|_{r=R_{o}} = 0. 
\label{eq1}
\end{equation}
This holds good because at the orbiting radius $r=R_{o}$, there is a classical turning point located at the maximum potential. Here, $V_{nucl}(r)$ is the nuclear potential, $V_{coul}(r)$ is the Coulomb potential, and $V_{cent}(r)$ is the centrifugal potential due to the orbital angular momentum of the dinuclear system. The nuclear orbiting phenomenon has been observed in the past for both elastic and inelastic scattering \cite{Braun}.
However, experiments have been carried out only at energies much higher than the fusion barrier energy \cite{Ray2007}, where the energy difference between the beam energy and fusion barrier is converted into internal excitation and rotation energy \cite{Shapira}. We will discuss what happens to the nuclear orbiting phenomenon if the experiment is conducted around the fusion barrier in following section. 
\par
\begin{table}[ht]
\textcolor{black}{\caption{Coupled-channel (CC) calculations for different systems necessitate the utilisation of potential parameters. The symbols, V$_0$, r$_0$, and a$_0$, stand for the potential depth, radius parameter, and diffuseness parameter for the real part of the interaction potential, respectively. While, the symbols W$_0$, r$_i$, and a$_i$ represent the potential depth, radius parameter, and diffuseness parameter, respectively, for the imaginary part of the interaction potential.}
\centering
\label{tab:potential_parameter}
\begin{tabular}{llllllll}
\hline
System             & V$_0$ & r$_0$ & a$_0$ & W$_0$ & r$_i$ & a$_i$ & Ref. \\ \hline 
$^{56}$Fe+$^{12}$C & 48.0  & 1.193 & 0.643 & 12.0  & 1.193 & 0.634 & \cite{roy2016measurement}    \\
$^{58}$Ni+$^{12}$C & 48.0  & 1.04  & 0.63  & 12.0  & 1.14  & 0.50  &   \cite{roy2016measurement}  \\
$^{63}$Cu+$^{12}$C & 40.0  & 1.24  & 0.61  & 24.0  & 1.25  & 0.56  & \cite{von1982inelastic}\\
\hline \\
\end{tabular}}
\end{table}
Comparison of the inelastic cross-sections shown in Fig.\ref{Fresco} with the nuclear orbiting cross-sections shown in Fig.\ref{orbiting} shows that the inelastic cross section curves show an increasing trend with the beam energies, while a resonance type of character is seen in the nuclear orbiting cross-section curves. Note that the potentials of Table II display resonances in the orbiting phenomenon at energies within the range considered in the experiments \cite{Prashant-PRL-2017}. However, these potentials do not show up resonances in the inelastic cross sections at these energies or at the small angular momenta due to the contribution of the other (non-resonant) partial waves. Moreover, the theoretical orbiting resonance energies are in excellent agreement with the measured values of $B_{res}$ \cite{Prashant-PRL-2017}. This undoubtedly establishes that the phenomenon of dinuclear orbiting resonance is responsible for the observed resonance in the x-ray spectroscopy experiment of ref \cite{Prashant-PRL-2017}. Besides, it also proves: (i) the phenomenon occurs below the fusion barrier but above the interaction barrier and (ii) the dinuclear orbiting complex influences on the atomic process.
\begin{table}[ht]
\caption {The excited states ($\lambda^{\Pi}$), excitation energies (E$_{\lambda}$), and deformation parameters ($\beta_{\lambda}$) for $^{56}$Fe, $^{58}$Ni, $^{63}$Cu 
and $^{12}$C nuclei were employed in the CC calculations.}
\label{tab:BE-2} 
\centering
\begin{tabular}{llllll}
\hline 
Nucleus & $\lambda^{\Pi}$~~~~ & E$_{\lambda}$ (MeV) & $\beta_{\lambda}$~~~~ & Ref. \\ \hline 
$^{56}$Fe & 2$^{+}$ & 0.846 & 0.2392 & \cite{raman2001transition} \\
$^{58}$Ni & 2$^{+}$ & 1.454 & 0.18 & \cite{raman2001transition} \\
$^{63}$Cu & 2$^{+}$ & 0.9 & 0.22 & \cite{von1982inelastic} \\ 
\textcolor{black}{$^{12}$C} &\textcolor{black}{2$^{+}$} & \textcolor{black}{4.44} & \textcolor{black}{-0.519} & \textcolor{black}{\cite{Newton2001}} \\ \hline 
\end{tabular}
\end{table}
\section{Inelastic cross section calculation}
\begin{figure}[!ht]
    \center
    \includegraphics[width=0.8\linewidth]{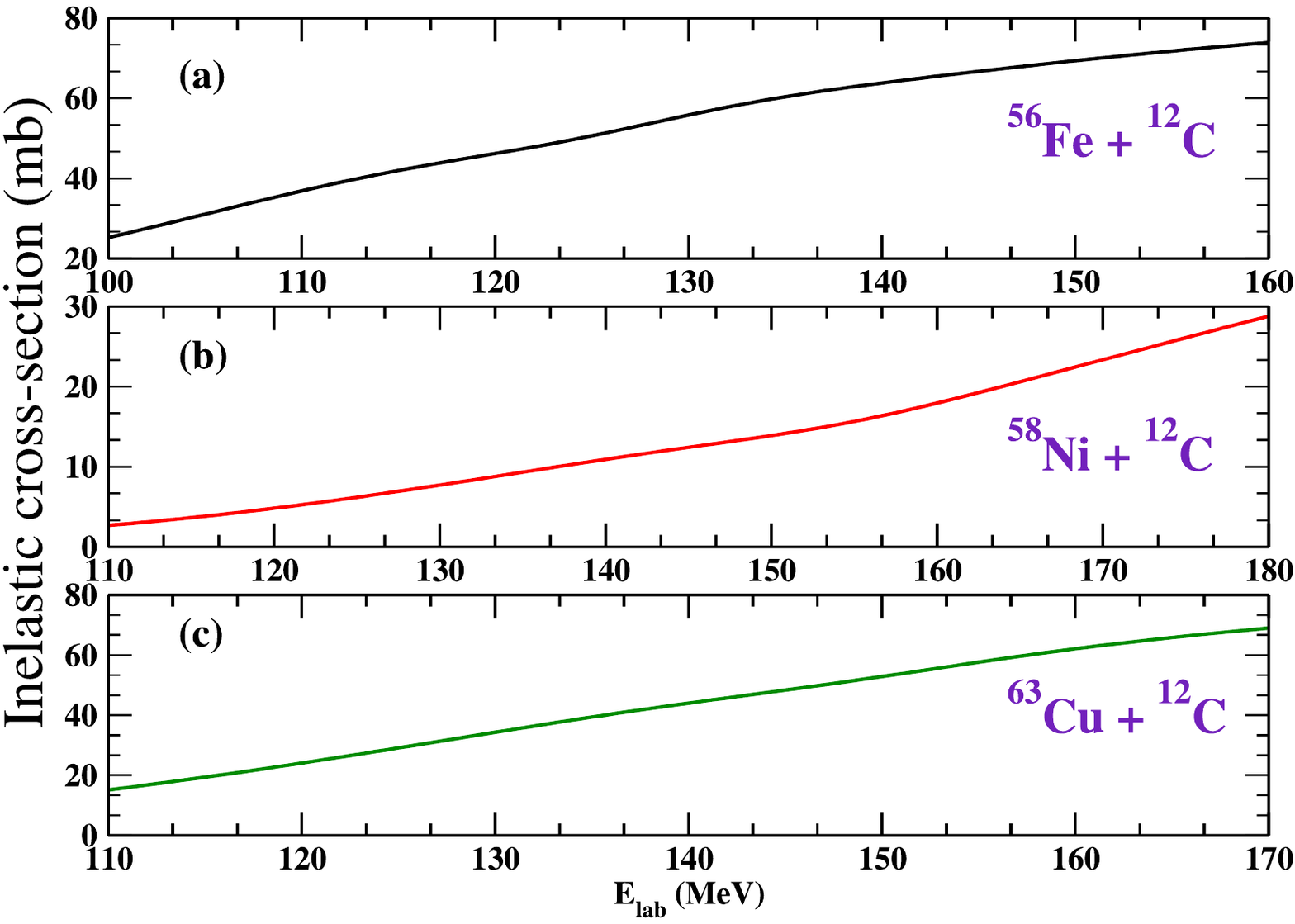}
    \caption{Inelastic scattering cross section as a function of lab energy for  the reactions $^{56}$Fe + $^{12}$C, $^{58}$Ni + $^{12}$C and $^{63}$Cu + $^{12}$C.}
    \label{Fresco}
\end{figure}

\textcolor{black}{Inelastic cross section for all three reactions $^{12}$C + $^{56}$Fe, $^{12}$C + $^{58}$Ni and  $^{12}$C + $^{63}$Cu has been calculated using the coupled-channel scattering (CCFULL-SC) code \cite{hagino2012subbarrier}. In the CCFULL-SC calculations, the total interaction potential is given below :}
\textcolor{black}{
\begin{equation}
\label{eq5}
V(R)= V_C(R) + V_N(R)+ V_{l}(R)
\end{equation}
}
\textcolor{black}{
Where V$_C$(R) is the Coulomb potential between the projectile and target nuclei:
\begin{equation}
V_C(R) = \begin{cases} 
\frac{Z_pZ_te^2}{R} &  R \geq R_c \\
\frac{Z_pZ_te^2}{2R_c}\left(3-\frac{R^2}{R_c^2}\right) &  R \leq R_c
\end{cases}
\end{equation}
Here R$_C$ = R$_1$ + R$_2$ is the sum of the radius of two nuclei given by R$_i$ = r$_c$ A$_{i}^{1/4}$, i = 1,2. $A_1$ (Z$_p$) and $A_2$ (Z$_t$) are the mass number (atomic number) of the projectile and target, respectively. The reduced Coulomb radius r$_c$ were taken as 1.1 fm. V$_N$(R) denotes nuclear potential in the Woods-Saxon form as below:}
\textcolor{black}{
\begin{equation}
V_N(R) = -\left[\frac{V_0}{1+\exp \left(\frac{R-(R_1+R_2)}{a_0}\right)}  + \frac{iW_0}{1+\exp \left(\frac{R-(R_1+R_2)}{a_i}\right)}\right].
\end{equation}}
\noindent The first term is a Woods-Saxon form of optical model nuclear potential in which the real parameters are the depth $V_0$, radius $r_0$ and diffuseness $a_0$. Whereas $W_0$, $r_i$ and $a_i$ in the second term are the imaginary depth of the potential, radius and diffuseness parameters, respectively. Both the real and imaginary parts of the potential parameters used in the calculations for the different reactions are given in Table \ref{tab:potential_parameter}.
The repulsive centrifugal potential V$_l$(R) is given as 
\begin{equation}
V_l(R) =  \frac{\ell(\ell+1)\hbar^2}{2\mu R^2}
\end{equation} 
Where $\ell$ denotes the angular momentum and $\mu$ the reduced mass. In these calculations, we considered the coupling to the one-phonon (2$^+$) vibrational state of the projectile \textcolor{black} {as well as the target nuclei in the present calculations}. The excited states ($\lambda^{\Pi}$), excitation energies (E$_{\lambda}$), and deformation parameters ($\beta_{\lambda}$) for $^{56}$Fe, $^{58}$Ni, $^{63}$Cu and $^{12}$C nuclei employed in the CC calculations have also been listed in Table III. The inelastic cross sections so obtained have been given in Fig. \ref{Fresco}.
%
\section{Dinuclear orbiting resonances around the fusion barrier: its time scale and cross section}
We now discuss in detail how the dinuclear orbiting resonances can occur in the sub-barrier energy regime. First of all, the projectile energy must be such that it can overcome at least the $B_{int}$ in order to experience the strong nuclear force. Next, the projectile ions with energy higher than $B_{int}$ must approach the target nucleus with a suitable range of impact parameters such that the distance of closest approach is less than the interaction barrier distance. As soon as these criteria are fulfilled, the projectile ion is under the influence of the nuclear interaction, and thus at that moment, it is pulled considerably towards the target nucleus. If the former has had sufficient energy, then the excess energy can be used up by several physical processes. However, the energy available in the studied systems  $^{12}$C($^{56}$Fe,$^{56}$Fe), $^{12}$C($^{58}$Ni,$^{58}$Ni) and $^{12}$C($^{63}$Cu,$^{63}$Cu) \cite{Prashant-PRL-2017} is only up to 3 - 4 MeV in the centre of mass frame, which is not sufficient even for a single neutron transfer, as discussed earlier. Furthermore, the charged particle transfer will change the projectile ion to a projectile like ion and that is not the case in the observed x-ray spectra. \\ 
\indent As an alternative, we show here that the dinuclear orbiting process is suitable to explain the origin of the resonance structure. In this process, the energy  difference between the beam energy ({\textit{E}}) and interaction barrier energy will be used up in the dinuclear orbiting with a certain orbital angular momentum. We can equate the sum of $V_{nucl}(r)$ and $V_{coul}(r)$ to the $B_{int}$, and the total potential energy $(V_{nucl}(r)+V_{coul}(r)+V_{cent}(r)$) at $r=R_{o}$  to the $B_{res}$. Thus, we obtain the centrifugal energy at $E=B_{res}$ as follows:
\begin{equation}
B_{res}-B_{int} = \frac{{L^2}{\hbar^2}}{2\mu{R_o^2}},
\label{eq2}
\end{equation}
Here, $L\hbar$ is the orbital angular momentum of the dinuclear complex and $\mu$ is the reduced mass of the projectile and target nuclei. We have obtained the values of $R_o$ and $L$ by solving Eqs. \ref{eq1} and \ref{eq2} for each dinuclear system,  which are given in Table \ref{first}. It can be seen that the $L$-value (rounded off) is 12$\hbar$ for the systems $^{56}$Fe + $^{12}$C and $^{58}$Ni + $^{12}$C, and 11$\hbar$ for the system $^{63}$Cu + $^{12}$C at the resonance condition. Though the value of the orbiting radius is somewhat smaller than the fusion barrier radius ($R_b - R_o \approx 0.2$ {fm}) for all the three systems, fusion does not occur because of the centrifugal force. 
\begin{table}
\caption{The lifetime of the dinuclear complex calculated from the present model $(t_m)$ and the nucleon exchange code HICOL $(t_h)$ \cite{HICOL} discussed above. Here, $E_x$, $\bigtriangleup E_x$, $\bigtriangleup Z_{p,eff}$ and $t_0$ are x-ray energy (keV) at the $B_{res}$, the measured jump in the x-ray energies (keV) \cite{Prashant-PRL-2017}, $C\times Z_t$, and characteristic time (see text), respectively.}
\centering
\begin{tabular}{c c c c c c c c }
\hline 
$Reaction$&$E_x$& $\bigtriangleup E_x$&$\bigtriangleup Z_{p,eff}$ & $C$&$t_0(as)$&$t_m(zs)$&$t_{h}(zs)$  \\ \hline
$^{12}C$+$^{56}Fe$ & $6.60$  & $0.025$ & $0.0473$ & $0.0079$& $0.383$ & $3.0$& $1.3$\\
 $^{12}C$+$^{58}Ni$ & $7.67$  & $0.032$ & $0.0563$ & $0.0094$ &$0.328$&$3.1 $& $1.2$\\
 $^{12}C$+$^{63}Cu$ & $8.30$  & $0.035$ & $0.0690$ & $0.0098$&$0.306$&$3.0$& $1.2$\\
\hline
\label{sec}
\end{tabular}
\end{table}
When the dinuclear complex is formed at $B_{res}$, the projectile and target nuclei are very close to each other at a distance $R_o$, while the orbital electrons in the projectile or target atoms  are far apart. Hence, the electrons of the projectile ion feel a two-center nuclear charge system due to the proximity of the projectile and target nuclei. However, this two-center system cannot be treated as one leading to molecular orbital x-ray process, which is characterised by continuous x-rays as opposed to a discrete x-ray peak is observed [2] and its centroid jumps as bombarding energy exceeds the $B_{int}$. Thereby, the effective Coulomb potential for a projectile electron at a distance ${r}$ from the projectile nucleus can be written as 
\begin{equation}
\label{eq3}
 V_{eff}(r) =\frac{-(Z_{p,eff} + C\times Z_t)\times e^2}{4\pi \epsilon_0  r}, 
\end{equation}
\noindent Here, $Z_{p,eff}$ is effective atomic number of projectile (($Z_p-s$); $s$ is screening constant for a multielectron system) and  $Z_t$ is atomic number of target. $C$ is a number $0 \le C \le 1$; its value depends on the lifetime of the dinuclear complex. According to Eq. (\ref{eq3}), the projectile x-ray energy increases with $C$. Hence, the experimental jump of the x-ray energy ($\bigtriangleup E_x$ in Fig. 2 of  ref.  \cite{Prashant-PRL-2017}) can be scaled up by $(Z_{p,eff}+C\times Z_t)^2$ to obtain the value of $C$ for the corresponding system. Next, we postulate that the value of $C$ varies with time, such that 
\begin{equation}
\label{eq4}
C = \overline{C(t)} =\frac{1}{t_0}\int_{0}^{t_f}f(t)dt= 1- \exp (-t_f/t_0).
\end{equation}
\noindent Here, $t_f$ is the time spent by the two nuclei together in the dinuclear complex and $f(t)=\exp (-t/t_0)$, where $t_0$  is characteristic time for the corresponding atomic state of the exiting projectile ion that is defined as the ratio of the expectation values of electronic radius ($\langle{r}\rangle$) and velocity ($\langle{v}\rangle$) of the state. According to the measured x-ray spectra \cite{Prashant-PRL-2017}, in the exit channel Li-like $1s2s2p$~$^{2,4}P^o_{1/2,3/2,5/2}$ levels ($s=1.2$; according to Slater's rule) in Fe, Ni and Cu are mostly populated at the resonance energies. To evaluate $t_0$ for these levels, we have evaluated $\langle{r}\rangle$ and $\langle{v}\rangle$ by the multi-configuration Dirac-Fock formalism using GRASP2K \cite{GRASP2K} atomic structure package. In this formalism the atomic state function, $\langle\Gamma{PJM_J}|$, for a given level is expressed as linear combination of configuration state functions as follows
\begin{equation}
\label{asf}
|\Gamma{PJM_J}\rangle=\sum_iC_i|\gamma_iPJM_J\rangle.
\end{equation}
Here, $\Gamma$ is the configuration of the atomic state $|\Gamma{PJM_J}\rangle$ and $\gamma_i$ is the configuration of the ith configuration state function $|\gamma_i{PJM_J}\rangle$. The symbols $P, J, M_J$ stand for parity, total angular momentum and magnetic quantum number of the atomic or configuration  state function. The configuration state functions are built from the sum of products of one electron Dirac orbitals. The angular and spin-dependent parts of these orbitals are assumed to be known, whereas the radial parts are determined by optimization procedure. In fact, the atomic state functions are optimized starting from the Dirac Hamiltonian in a self-consistent field procedure, where both the radial parts of the one-electron Dirac orbitals and the expansion coefficients $C_i$s are optimized.  In the GRASP2K package all calculations are done with non-interacting blocks of given parity and $J$ value.  For further details, see \cite{GRASP2K}. The expectation value of radius $\langle r^k\rangle$ and kinetic energy $\langle{T}\rangle$ are computed as follows
\begin{equation}
\label{eqrk}
\langle{r^k}\rangle=\langle\Gamma PJM_J|\sum_{i=1}^N[r^k_i]|\Gamma PJM_J\rangle
\end{equation}
and
\begin{equation}
\label{eqke}
\langle{T}\rangle=\langle\Gamma PJM_J|\sum_{i=1}^N[c\boldsymbol\alpha_i\mathbf{.{p}_i} + ({\boldsymbol\beta}_i-1)mc^2]|\Gamma PJM_J\rangle
\end{equation}
Where $c\boldsymbol{\alpha.{p}} + {\boldsymbol\beta}{mc^2}$ is the Dirac Hamiltonian and $c\boldsymbol{\alpha.{p}} + (\boldsymbol{\beta} -1){mc^2}$ is Dirac kinetic energy operators. All the symbols here have their usual significance. $\langle{v}\rangle$ is evaluated from $\langle{T}\rangle$ using the relation
\begin{equation}
\label{eqke1}
\frac{\langle{v}\rangle^2}{c^2}=1-\left(1+\frac{\langle{T}\rangle}{m c^2}\right)^{-2}.
\end{equation}

The above relation is obtained from the relativistic kinetic energy relation $\langle{T}\rangle={(p^2c^2+m^2c^4)}^{1/2}-mc^2$. Here, $p$ is the momentum of the orbital electron, $c$ is the velocity of light and m is the rest mass of the electron.
\par
The value of $t_0$ for the three systems is given in Table \ref{sec}. Eq. (\ref{eq4}) suggests that $C(t_f=0)=0$ and $C(t_f=\infty)=1$. The measured value of $C$ can be used to determine the lifetime of the dinuclear orbiting complex from Eq.(\ref{eq4}), giving rise to the value of $t_m$ listed in Table \ref{sec}. The lifetime of the dinuclear complex can also be estimated by a one-body dissipation model using the HICOL code \cite{HICOL}. The dynamical HICOL model treats colliding nuclei as the spheres of Fermi gas and permits exchange of particles, momentum, and entropy through a small contact area in the mean single-particle potential. Time evolution in the colliding trajectories is estimated by solving a Langevin equation with a fluctuating dissipative force. The behavior of fluctuating force is understood in terms of microscopic pictures of particle exchange between two nuclei. The model assumes two spheres are connected by a neck and their dynamical evolution described by a sequence of shapes keeping the mass and charged density unchanged during the collision to conserve the volume of shape. Further details can be seen in \cite{HICOL,patil2020pre}.
\par
One can see in Table \ref{sec} that the lifetime of the  dinuclear orbiting complex obtained from our model, $t_m$ (=$t_f$), for the systems used are about a factor of 2.5 larger than that of the prediction, $t_h$, of the one-body dissipation model using the nucleon exchange code HICOL \cite{HICOL}. The latter works only at energies greater than the fusion barrier so that nucleon exchange may take place in a non-equilibrium condition between friction and diffusion. Thus, $t_h$ gives a measure of the time taken to come from a statistically non-equilibrium to a statistically equilibrium state by particle exchange process. In the present case particle exchange can be possible only through neutrons or $\gamma$-photon as the phenomenon being studied through the projectile x-ray emissions.  The HICOL model does not work below the Coulomb barrier energies. We have used just above the Coulomb barrier energies (2.3, 2.4 and 2.5 MeV/u for $^{56}$Fe, $^{58}$Ni and $^{63}$Cu experiments, respectively). These energies are higher than the experimental resonance energies (2.14, 2.31 and 2.27 MeV/u for $^{56}$Fe, $^{58}$Ni and $^{63}$Cu experiments, respectively). The main reason for the difference between the present model and HICOL model may be due to the difference between the resonance energies observed and energies used for HICOL calculations. Though the difference is quite small, it still happens because the events below the barrier are strongly damped and that may enhance the life of the dinuclear system. Furthermore, the nuclear orbiting process in the above barrier region also is quite slower than dissipative nucleon exchange process as has been pointed out quite sometime back \cite{ray2007nuclear}.  Hence, this comparison suggests that the nuclear orbiting model is a better representation than the dissipative particle exchange model to explain the resonance structure observed in the experiment \cite{Prashant-PRL-2017}.
\par
\begin{figure}
  \centering
    \includegraphics[width=0.9\linewidth]{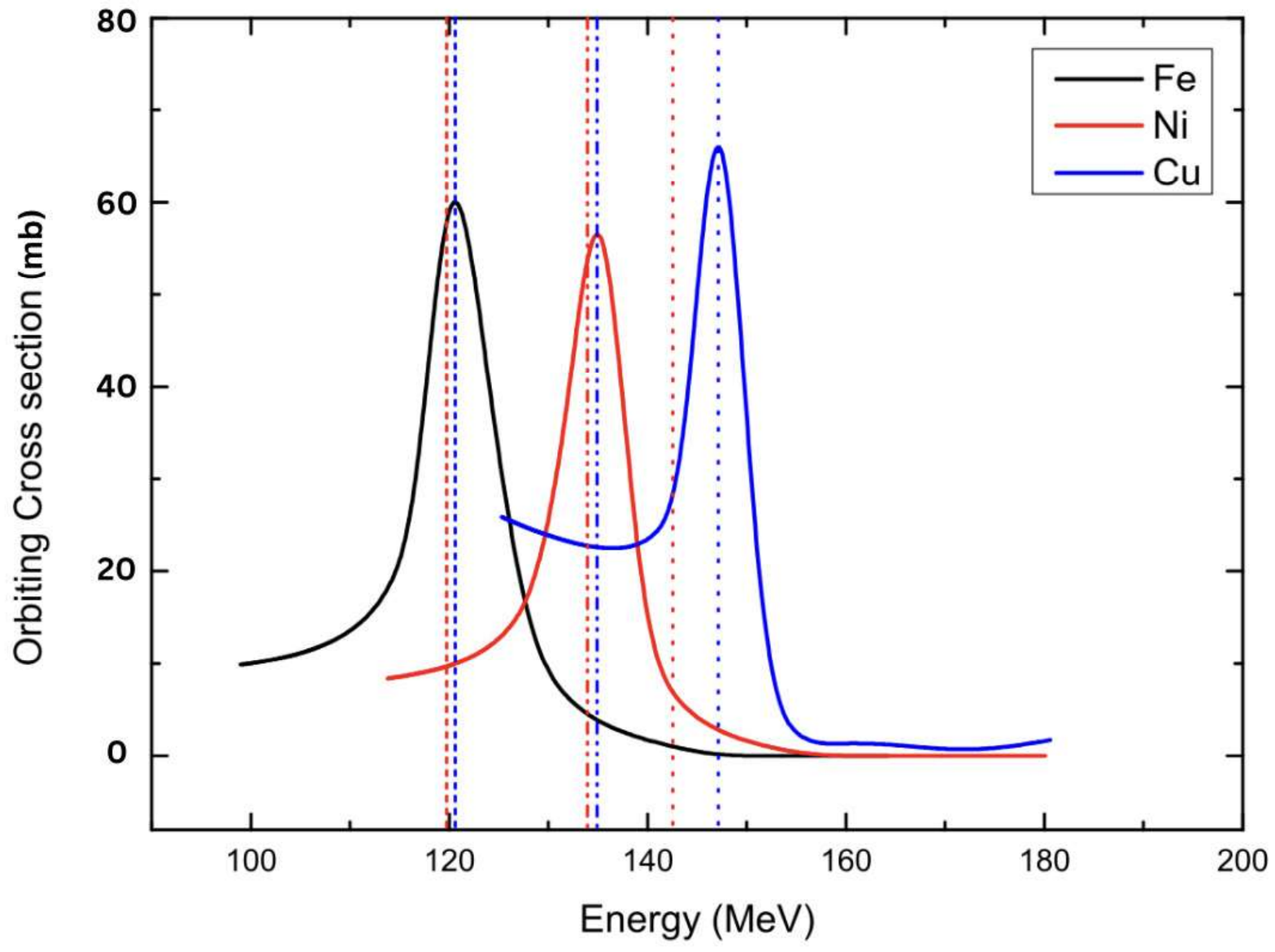}
    \caption{Nuclear orbiting  cross-section (solid lines) as  a  function  of  lab energy for $^{56}$Fe +$^{12}$C,$^{58}$Ni + $^{12}$C and $^{63}$Cu + $^{12}$C dinuclear systems. Red vertical line corresponds to the measured resonance energy ($B_{res}$) \cite{Prashant-PRL-2017} and blue vertical line indicates the theoretical resonance energy using the scattering theory \cite{Korsch_1983}. This theory does not consider the nuclear structure effect in the calculation and thus, a difference  is seen between the measured and theoretical resonance energies for $^{63}$Cu + $^{12}$C system due to an unpaired nucleon in $^{63}$Cu. }
    \label{orbiting}
\end{figure}
\par
We estimate now the orbiting cross-section of the dinuclear complex with the prescription of ref. \cite{Korsch_1983}. In this theory, a particle with a reduced mass $\mu$ is orbiting at an internuclear distance $r$ \textcolor{black}{in a Woods-Saxon form of inter-nucleus potential V(r) given in Eq. \ref{eq5}. Next, we compare the orbiting cross-sections so obtained with the inelastic cross-sections for all three reactions $^{12}$C + $^{56}$Fe, $^{12}$C + $^{58}$Ni and  $^{12}$C + $^{63}$Cu. We note that the Woods-Saxon form of the inter-nucleus potential given in Eq. \ref{eq5} and the corresponding potential parameters given in Table \ref{tab:potential_parameter} are also used for the nuclear orbiting cross section calculations. Comparison of the inelastic cross-sections shown in Fig.\ref{Fresco} with the nuclear orbiting cross-sections shown in Fig.\ref{orbiting} shows that the inelastic cross section curves show an increasing trend with the beam energies, while a resonance type of character is seen in the nuclear orbiting cross-section curves. It suggests that the potentials of Table II display resonances)(Fig.\ref{orbiting}) in the orbiting phenomenon at energies within the range considered in the experiments \cite{Prashant-PRL-2017}. However, these potentials do not show up resonances in the inelastic cross sections at these energies (Fig.\ref{Fresco}) or at the small angular momenta due to the contribution of the other (non-resonant) partial waves. Moreover, the theoretical orbiting resonance energies are in excellent agreement with the measured values of $B_{res}$ \cite{Prashant-PRL-2017}. This undoubtedly establishes that the phenomenon of dinuclear orbiting resonance is responsible for the observed resonance in the x-ray spectroscopy experiment of ref \cite{Prashant-PRL-2017}. Besides, it also proves: (i) the phenomenon occurs below the fusion barrier but above the interaction barrier and (ii) the dinuclear orbiting complex influences the atomic process.}
This inference is in support of the fact discussed above that the orbiting forms around the fusion barrier radius where fusion is hardly possible, at low projectile energy and with small angular momentum (see Eq.(9-11)). Even at high energies, the large cross sections are measured for an orbiting, dinuclear configuration \cite{bhattacharya2005survival}.
\par
From conservation of energy, $E$, and angular momentum, $L$, the net deflection angle as a function of $L$ of a particle can be estimated from ref.  \cite{Korsch_1983} as
\begin{equation}
\label{eq6}
\Theta (L) = \pi - 2\int_{r_i}^{\infty} \frac{L}{r^2}[2\mu (E - V(r)) - L^2/r^2]^{-1/2}dr.  
\end{equation}
Here, the lower limit of the integration, $r_i$, is the outermost turning point, which is taken as the orbiting radius, $R_0$. The orbiting region is restricted by the rainbow value [18] of angular momentum ($L_R$).  For example, for $^{12}$C + $^{56}$Fe scattering $L_R$ $\approx$ 25$\hbar$, which restricts the orbiting region within $L < 25 \hbar$. Further, in the experiment the effect of individual trajectories is not seen and one observes the total contribution of all the trajectories.
In order to get the differential cross section, we need to integrate $I_j(\vartheta)$, single orbit contribution with a particular value of angular momentum ($L=L_j$),  over all values of $\vartheta$ and $L$ for an interval where the orbiting occurs:
\begin{equation}
\frac{d\sigma}{d\Omega} = \int \int I_j(\vartheta) d\vartheta dL~~{\rm fm^2}, 
\label{eq8}
\end{equation}
where $I_j(\vartheta)$ is given by the relation [24]
\begin{equation}
I_j(\vartheta) = \bigg| \frac{1}{2\mu E \sin \Theta (d\Theta /dL)} \bigg|_{L=L_j} \frac{\rm fm^2}{\rm MeV~s}.
\label{eq7}
\end{equation}
 Here, the units used are as follows: radius in $fm$, mass in $MeV~c^{-2}$, energy in MeV, and angular momentum in ${\rm MeV~s}$. The scattering angle $\vartheta$ is related to the deflection function $\Theta$ as
 \begin{equation}
\Theta(L_j)= \pm{\vartheta}-2\pi n_j ~~~~~~~~~j=1,2,...
\label{eq9}
\end{equation}
 where $n_j$ is the number of rotations of dinuclear complex around the scattering centre. The $L_j$ value is taken from Eq.(\ref{eq2}) of this work in which $B_{res}$ is replaced by {\textit{E}} and $E > B_{int}$. Differential cross section given in Eq.(\ref{eq8}) here is a function of {\textit{E}}, we get the cross section for the nuclear orbiting complex as a function of energy as shown in Fig.\ref{orbiting}.
 \begin{figure}[!ht]
  \centering
    \includegraphics[width=0.9\linewidth]{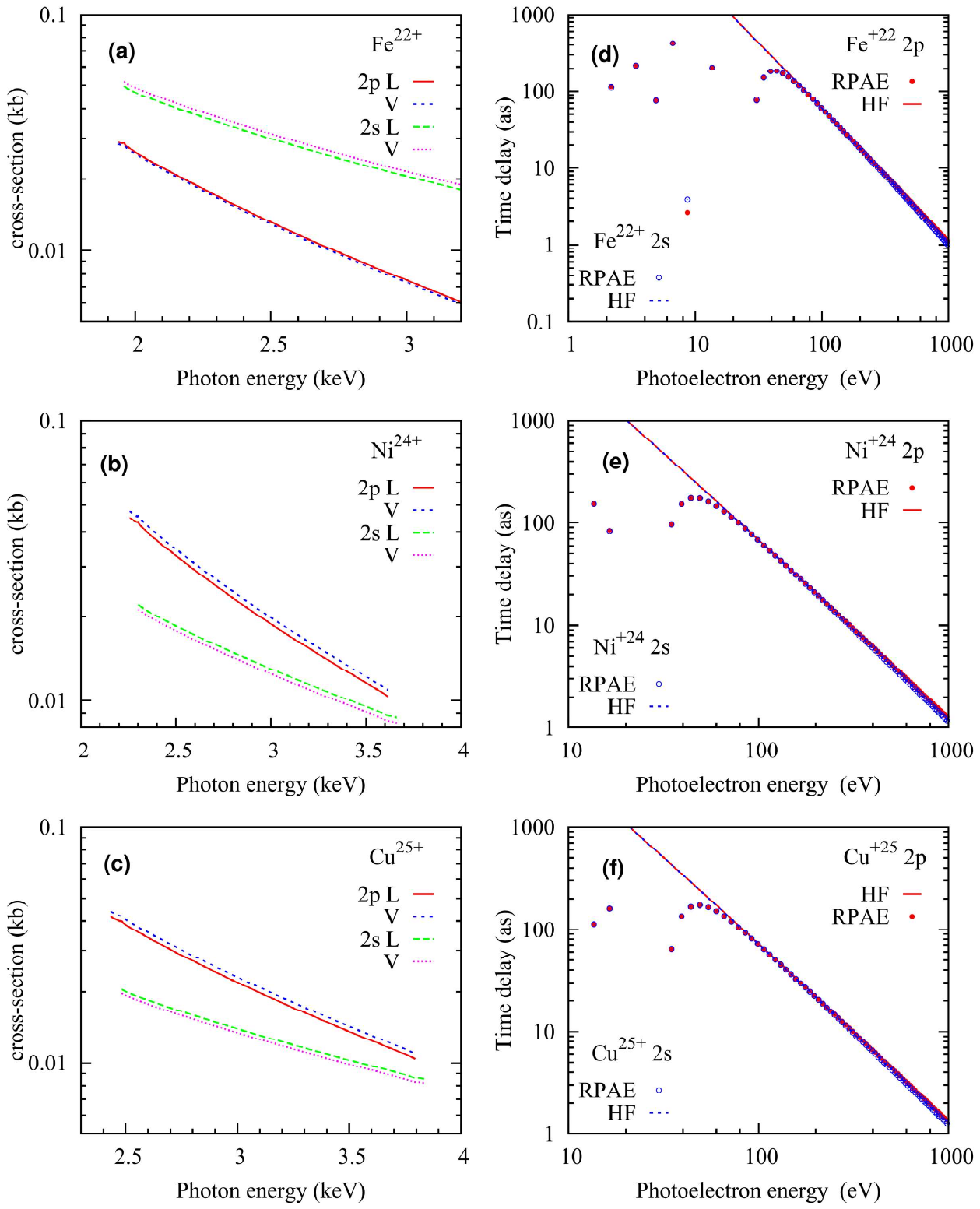}
  \caption{Photoionization cross-section as a function of photon energy of the emitted photons during the dinuclear orbiting is shown in the left panel ((a), (b), and (c)).  The Wigner time delay due to orbiting induced photoionization versus photoelectron energy is shown in the right panel ((d), (e), and (f)). }
  \label{fig:delay}
\end{figure} 
The nuclear orbiting cross-section is large, but it lasts only for a few $zs$. This fact indicates that a disruptive force restricts the nuclear orbiting to survive for a long time. This disruption is nothing but the photon/multiphoton exchange between the two nuclei during the orbiting, which results in both the projectile and target nuclei excited. Thus, the Eq. (\ref{eq2}) can be rewritten as
\begin{equation}
(E-\Delta E + \Delta E^\prime) - (B_{int} + \Delta E - \Delta E^\prime) = \frac{{L^\prime}^2 \hbar^2}{2\mu{R_o^2}}.
\label{eq19}
\end{equation}
Here, $B_{res}$ is replaced by the beam energy, $E$, which is greater than $B_{int}$.  $\Delta E$ is the energy flowing from the projectile to target,  $\Delta E^\prime$ is the energy received by the projectile from the target due to a single multiphoton process \cite{PRL1987} (more specifically, multiphoton exchange via a single virtual photon), and $L^{\prime}$ is the angular momentum resulted from the redistribution of the energies. If the change in angular momentum is high enough, the dinuclear complex will break into its constituents. This gives us a qualitative understanding about the phenomenon.
\section{Wigner-Smith time delay}
The above multiphoton or even a single photon exchange can be accountable in an atomic excitation as well. This excitation in an ion with vacancy in the K-shell can lead to an additional autoionization different from what happens normally owing to only the K-shell vacancy before x-ray transition takes place and the K-vacancy is filled. This process is likely to result in higher charge state on the ion having K-vacancy and thus higher x-ray energy, as has been observed \cite{Prashant-PRL-2017}. The autoionization process does not occur instantly as the electron has to move from the interaction regime to an interaction free zone, which introduces a time delay because of the difference between the density of states of the two regions \cite{ahmed2004number}. The energy-integral of time delay is an adiabatic invariant in quantum scattering theory and it provides a quantization condition for resonances \cite{jain2005}. The delay in the autoionization process, called as the Wigner-Smith time delay \cite{wigner55,smith60}, has been measured in an experiment \cite{Schultze2010}. To estimate the Wigner-Smith time delay as well as the photoionization cross-section, we have employed the non relativistic versions of the random-phase approximation with exchange (RPAE) and Hartree-Fock (HF) methods \cite{ASK2015}. A brief description of RPAE model is given bellow. 
RPAE has been a well established and extensively tested technique. It has a long track record of successful applications to calculations of total and partial photoionization cross-sections of valence \cite{A90} and inner shells \cite{PhysRevLett.85.4703} of noble gas atoms. Its more recent applications can also be found  \cite{PhysRevA.87.063404,PhysRevA.92.063422}. Valence shell calculations of time delay have been tested in numerous experiments \cite{0953-4075-47-24-245003,0953-4075-47-24-245602,Jain:18,Jain:18a}.

 Here we follow closely the photoionization formalism as outlined in \cite{A90}.  We evaluate the single-photon dipole matrix element
$\langle \psi^{(-)}_\textbf{k}|\textbf{\^{z}}| \phi_i \rangle$
 from a bound state 
$ \phi_i(\textbf{r}) = Y_{l_im_i}(\textbf{\^{r}})R_{n_il_i}(r) $
to an incoming scattering state with the given photoelectron momentum $\k$: 
\begin{eqnarray}
\label{partial} 
\psi^{(-)}_{\k}(\r) = \frac{(2\pi)^{3/2}}{k^{1/2}}\hs{-1mm}\displaystyle\sum_{lm} i^le^{-i\delta_l(E)}Y^*_{lm}(\hat{\k})Y_{lm}(\hat\r)R_{kl}(r) \ .
\end{eqnarray}
We conduct the spherical integration to arrive to the following expression:
\ba
\label{dipole} 
\langle \psi^{(-)}_\k|\hat z| \phi_i\rangle &=&  \frac{(2\pi)^{3/2}}{k^{1/2}}\sum_{l=l_i\pm1, m=m_i}e^{i\delta_l(E)}i^{-l}Y_{lm}(\hat \k)\, \\&&\hs{5mm}\times\left(\begin{array}{rrr} l&1&l_i\\ m&0&m_i\\
\end{array}\right) \langle kl \|\,\hat D\,\|n_il_i\rangle \nn 
\ea
Here $\langle kl \|\,\hat D\,\|n_il_i\rangle$ is the reduced dipole matrix
element, stripped of all the angular momentum projections.
The partial photoionization cross-section for the transition from an
occupied state $n_il_i$ to the photoelectron continuum state $kl$ is
calculated as
\be
\label{CS-HF} \sigma_{n_il_i\to kl}(\omega) = \frac43 \pi^2\alpha a_0^2\omega
\left| \langle k l\,\|\,\hat D\,\|n_il_i \rangle \right|^2 \ ,  
\ee 
$\omega$ being the photon energy, $\alpha$ the fine structure constant
and $a_0$ the Bohr radius Atomic units $e=m=\hbar=1$ are used in this
expression and throughout the paper.

In the independent electron Hartree-Fock (HF) approximation, the reduced dipole matrix element is evaluated as a radial integral,
\be
\label{reduced} 
\langle kl \|\,\hat D\,\|n_il_i\rangle = [ l][ l_i] \left(
\begin{array}{rrr} l&1&l_i\\ 0&0&0\\
\end{array} \right)\\ \int r^2dr \, R_{kl}(r)\,r\,R_{n_il_i}(r) \ , 
\ee
where the notation $[l] = \sqrt{2l+1}$ is used.
The basis of occupied atomic states $\|n_il_i\rangle$ is defined by the
self-consistent HF method and calculated using the computer code
\cite{CCR76}. The continuum electron orbitals $\langle kl \|$ are defined
within the frozen-core HF approximation and evaluated using the
computer code \cite{CCR79}. 

\begin{figure}[!t]
\includegraphics[width=1.0\columnwidth]{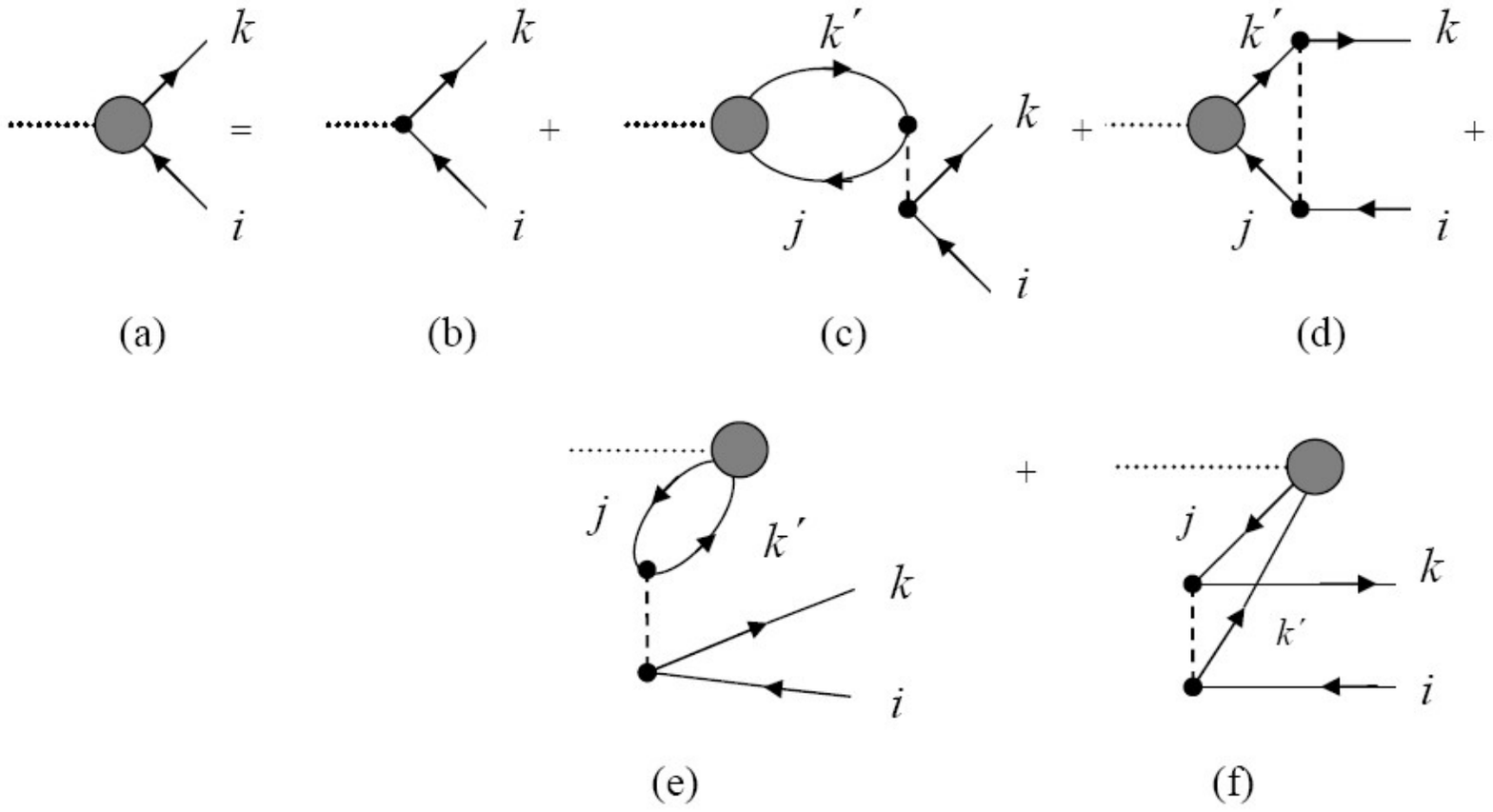}
\caption{Diagramatic representation of the photoionization amplitude
 $\langle kl|\hat{D}|n_il_i\rangle$ in the RPAE. Here, the time axis is directed from  left to right, the lines with arrows to the left (right) correspond to holes (electrons) in a filled atomic shell, a dotted line represents an incoming photon, a dashed line represents the Coulomb interaction between charged particles, and a shaded circle marks the effective operator $\hat{D}$ for the photon-atom interaction, which accounts for electron correlation in the atom.}
\label{diagram}
\end{figure}

In the random phase approximation with exchange (RPAE), the reduced
dipole matrix element is found by summing an infinite sequence of
Coulomb interactions between the photoelectron and the hole in the
ionized shell. This leads to a system of integral equations which can
be represented graphically by the diagrams as shown in Fig. \ref{diagram}.
There, diagram (a) represents the sum of all Coulomb interactions, diagram (b) depicts the HF term given by Eq.\ref{reduced} and diagrams
(c)--(f) represent RPAE corrections.  Diagrams (c) and (d) are known
as time direct (forward) and (e) and (f) as time reverse (backward). Diagrams (d) and (f)
account for the exchange interaction in the atom, thus being called
the exchange diagrams. As is seen from Fig. \ref{diagram}, a virtual excitation in the shell $j$ to the ionized electron state $\k'$ may
affect the final ionization channel from the shell $i$. This way RPAE
accounts for the effect of inter-shell $i\leftrightarrow j$
correlation, also known as inter-channel coupling.  It is important to
note that, within the RPAE framework, the reduced dipole matrix element is 
complex and, thereby, adds to the phase of the dipole amplitude.
\par
The photoelectron group delay, which is the energy derivative of the
phase of the complex photoionization amplitude, is evaluated as
\be
\label{delay}
\tau = \frac{d}{dE} \arg f(E)\equiv{\rm Im} \Big[ f'(E)/f(E) \Big] 
\ee
Here $f(E)$ is used as a shortcut for the amplitude $ \langle\psi^{(-)}_\k|\hat z| \phi_i\rangle $ given by \Eref{dipole} and evaluated
for $E=k^2/2$ and $\hat \k\parallel\z$.
\par
\begin{figure}
  \centering
    \includegraphics[width=0.6\linewidth]{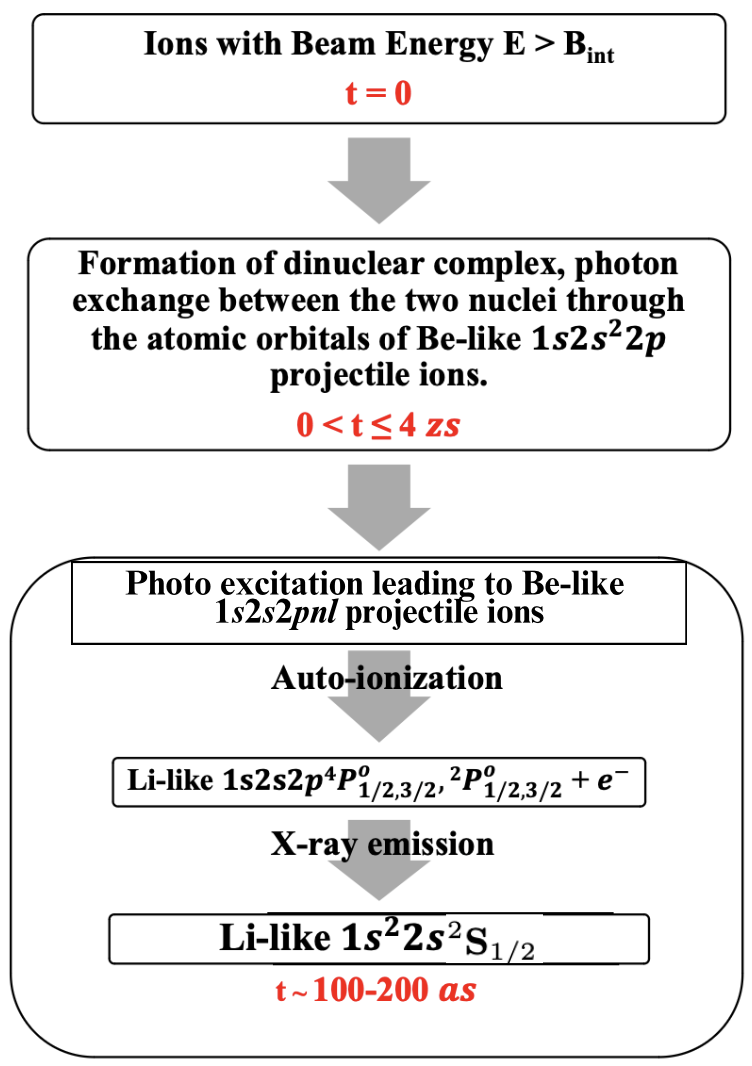}
  \caption{Schematic of the overall mechanism that reflects the influence of nuclear orbiting resonance on the observed atomic resonance in x-ray spectroscopy experiments \cite{Prashant-PRL-2017}.}
\label{schematic}
\end{figure}

For the photoionization and time delay calculations, we have considered charge species Fe$^{22+}$, Ni$^{24+}$, and Cu$^{25+}$ for the systems $^{56}$Fe + $^{12}$C, $^{58}$Ni + $^{12}$C, and $^{63}$Cu + $^{12}$C, respectively, as observed \cite{Prashant-PLA}. The Wigner-Smith time delay, $\tau$ ($\propto \varepsilon ^{- 3/2} ln (1/\varepsilon)$, where $\varepsilon$ is the photon energy), and   the elastic scattering phase, $\sigma$ ($\approx$ $\eta$ $ln {|\eta|}$, $\eta = -Z/\sqrt{2\varepsilon}$) are divergent near the threshold because of the Coulomb singularity \cite{ASK2015}. To remove this singularity, we have cut off the low energy photoelectrons for time delay calculations in present computations. The results of the computations are presented in  Fig.\ref{fig:delay}. The calculated values of  photoionization cross-sections are very close  for the length ({\textit{L}}) and velocity ({\textit{V}}) gauges. The photoionization cross-sections for 2p electron in Ni$^{24+}$ and Cu$^{25+}$ are higher than that of 2s electron for all photon energies, therefore these two systems will dominantly decay through the photoionization of 2p electron. Whereas, for Fe$^{22+}$, the  photoionization will dominate through 2s electron. Further, the Wigner-Smith time delay  can transfer the  orbiting  induced ionization  triggered  in the nuclear time scale ($zs$) to a few hundreds of $as$ for all the three systems considered here. Thereby, the photoionization phenomenon can occur in the atomic time scale and one measures the x-ray emissions. This entire mechanism is schematically shown in Fig.\ref{schematic}.\\
\indent A large  discrepancy between the measurements of the fission time scales using  the nuclear \cite{SHEN} and atomic techniques using blocking methods \cite{PhysRevLett.70.537} and x-ray fluorescence \cite{fregeau2012x} remains unresolved since 1990's. Even though a number of studies have been made to unfold this long standing issue, no convincing explanation is found until 2016 \cite{SIKDAR}. Though the anomaly between nuclear and blocking techniques has been resolved recently by Nandi $et~al.$ \cite{Nandi2022on} the discrepancies still exists between the nuclear and x-ray fluorescence techniques. Now this anomaly can also be removed with the above mentioned fact that a phenomenon occurring at the nuclear time scale, gets transferred to 
the atomic time scale via the Wigner-Smith time delay mechanism. The concerned study will be published elsewhere. 
\section{Summary and conclusion}
To summarise, we discussed about all the possible phenomena in section II which may cause an atomic resonance observed in the x-ray spectroscopy experiments and came to conclusion that the nuclear orbiting is the most probable process. The target and projectile nuclei in the dinuclear complex are so close that higher nuclear charge is felt by the orbital electrons. On the basis of this fact we have developed  a simple model, which enables us (i) to determine the orbiting duration for the dinuclear complex, which compares quite well with the predictions of the nucleon exchange code HICOL \cite{HICOL},  (ii) To draw a significant point is that the experimentally observed resonance energy in the x-ray spectra \cite{Prashant-PRL-2017} is well reproduced by the scattering theory of Korsch and Thylwe \cite{Korsch_1983}, (iii) To infer an interesting fact is that the interaction barrier radius is found to be a good measure of the range of the nuclear force. Inelastic cross section calculation using the coupled-channel scattering theory given in section III and  the orbiting cross-section of the dinuclear complex given in section IV reveals that the nuclear orbiting phenomenon is the physical process which leads to the an atomic resonance in the x-ray spectroscopy experiments. Descriptions given in section IV and V help us to succeed in unraveling the possible origin of the short time scale of the orbiting complex. It is governed by the inelastic excitation process involving multi-photon exchange between the projectile and target nuclei. The multi-photon exchange causes not only the inelastic excitation; but also the atomic excitation. This excitation in an atomic system with a vacancy in the  K-shell can lead to autoionization, which does not occur instantly; rather it goes through the Wigner-Smith time delay of the order of a few hundreds of $as$ (section V). Hence, the orbiting induced ionization triggered in the nuclear time scale ($zs$) transfers the phenomenon occurring in the atomic time scale ($as$) so that one can measure the x-ray emissions.  This crossover process between the different time scales may help resolve a longstanding  anomaly of fission time scales.
\par
To conclude, we have revealed that the nuclear orbiting resonance formed at the sub-barrier energies is observed as an atomic resonance in the x-ray spectroscopy experiments. This process explains every aspect of the observed resonance, such as the enhanced ionization, anomalous large-angle scattering and closeness of resonance to the fusion barrier, unlike the fact that the nuclear recoil induced shake off ionization \cite{Prashant-PRL-2017} describes well only the enhanced ionization. The present study will stir new interest in wider community crossing the borders of various disciplines between the molecular and solid state to the nuclear and high energy physics.
\section{Author Contributions Statement}
T.N. formulated the problem and wrote the main manuscript text, Y.K. used HICol code and did many other calculations, A.P.M. and G.S. used GRASP code, C.K. calculated inelastic cross sections and prepared Fig.1, N.R.D. ans S.R.J. calculated dinuclear orbiting cross section and prepared Fig.2, N.S.and H.C.M. attempted to calculate inelastic cross sections, A.S.K. calculated Wigner-Smith time delay and prepared Fig.3 and 4, T.N. summarized schematically the overall mechanism of nuclear orbiting resonance on the observed atomic resonance in x-ray spectroscopy experiments and prepared Fig.5. All authors reviewed the respective part of the manuscript.
\section{Data availability}
All data generated or analyzed during this study are included in this published article.
\begin{acknowledgments}
It is our pleasure to thank Prashant Sharma, Subir Nath, Amlan Ray, Arun Kumar Jain, Istiak Ahmed, Rakesh Kumar, R.P. Singh, R.K. Bhowmik, M. Dasgupta, and A. Chatterjee for illuminating discussions.

\end{acknowledgments}

\bibliographystyle{unsrt}
\bibliography{main}
\end{document}